\documentclass[aps,prd,amssymb,amsmath,amsfonts,superscriptaddress,nofootinbib,eqsecnum,reprint,showpacs,longbibliography]{revtex4-2}

\usepackage{amsmath}
\usepackage{amsfonts}
\usepackage{amssymb}	
\usepackage{graphicx}
\usepackage{bm}
\usepackage{color}
\usepackage{commath}

\usepackage{graphicx}
\usepackage{dcolumn}

\usepackage{color}
\usepackage{xcolor}
\definecolor{linkcolor}{rgb}{0.0,0.3,0.5}
\usepackage[unicode,colorlinks=true,citecolor=linkcolor,linkcolor=linkcolor,urlcolor=linkcolor]{hyperref}

\usepackage{cleveref}
\usepackage{orcidlink}

\graphicspath{{../figures/}}

\newcommand{\gammaAB}{\gamma_{\rm AB}}
\newcommand{\Sp}{\mathcal{S}_+}
\newcommand{\Sm}{\mathcal{S}_-}
\newcommand{\thp}{\theta_+}
\newcommand{\thm}{\theta_-}
\newcommand{\Thp}{\Theta_+}
\newcommand{\Thm}{\Theta_-}
\newcommand{\thav}{\langle\theta \rangle}
\newcommand{\Thav}{\langle\Theta \rangle}

\allowdisplaybreaks[1]
\newcommand{\Cambridge}{Department of Applied Mathematics and Theoretical Physics, University of Cambridge, Wilberforce Road CB3 0WA Cambridge, United Kingdom}
\newcommand{\Turin}{INFN Sezione di Torino, Via P. Giuria 1, 10125 Torino, Italy}

\begin{document}
	
	\title{Angular Momentum Flux in Scalar-Tensor Theories up to 1.5 post-Newtonian Order}.
	
	\author{Tamanna \surname{Jain}}
	\email{tj317@cam.ac.uk}
	\affiliation{\Cambridge}
	
	\author{Piero \surname{Rettegno} \orcidlink{0000-0001-8088-3517}}
	\email{piero.rettegno@to.infn.it}
	\affiliation{\Turin}
	
	\date{\today}
	
	\begin{abstract}
	We compute the angular momentum flux from a non-circular nonspinning binary system of compact objects in massless scalar-tensor theories up to one and a half post-Newtonian (1.5PN) order using multipole moments.  
	The angular momentum flux in scalar-tensor theories involves both a tensorial and a scalar contribution which can be further decomposed as instantaneous, tail and non-linear memory effects up to 1.5PN order. 
	We compute the explicit expressions of tail effects using the Fourier decomposition of tensorial and scalar multipole moments, and non-linear memory effects using the Newtonian order quasi-Keplerian representation of elliptic orbits in scalar-tensor theories. 
	This work is important to construct the radiation-reaction force and hence the waveform templates for eccentric binaries in scalar-tensor theories of gravity.
	\end{abstract} 
	
	\maketitle
	\section{Introduction}

	The detection of gravitational wave events by the LIGO~\cite{TheLIGOScientific:2014jea}, Virgo \cite{VIRGO:2014yos} and KAGRA~\cite{KAGRA:2020agh} interferometers, as well as future instruments like LISA, Einstein Telescope \cite{Maggiore:2019uih} and Cosmic Explorer \cite{Evans:2021gyd}, has opened the possibility of testing General Relativity in strong-field regime.

	The simplest extension of General Relativity (GR) to include additional fields is to consider scalar fields non-minimally coupled to gravity. 
	These ST (ST) theories have been extensively studied in the literature both observationally~\cite{Damour:1996ke, Freire:2012mg, Shao:2017gwu, Kramer:2021jcw, Zhao:2022vig, Gautam:2022cpb}, numerically~\cite{Healy:2011ef, Barausse:2012da, Berti:2013gfa, Shibata:2013pra, Palenzuela:2013hsa, Taniguchi:2014fqa}, and theoretically~\cite{Damour:1992we, Damour:1993hw, Mirshekari:2013vb, Zaglauer:1992bp, Will:2014kxa, Lang:2013fna, Lang:2014osa, Bernard:2018hta, Bernard:2018ivi, Brax:2021qqo, Shiralilou:2021mfl,
 Khalil:2022sii, Bernard:2022noq, Bernard:2023eul},  especially within the post-Newtonian (PN) formalism. 
	The equations-of-motion (EOM) and the mass quadrupole moment for the nonspinning two-body problem in ST theories are known up to 3PN order~\cite{Bernard:2018hta, Bernard:2018ivi} and 1.5PN order~\cite{Bernard:2022noq}, respectively. 
	This implies that both the energy flux and angular momentum flux up to 1.5PN order can be computed for inspiralling compact binaries moving in general \textit{non-circular} orbits. 
	The 1.5PN contributions to both energy and angular momentum fluxes in ST theories contain instantaneous terms as well as (non-linear) hereditary contributions \cite{Bernard:2022noq}.

	In this work, we use the results of Refs.~\cite{Lange:2018pyp, Lange:2017wki, Bernard:2022noq} and, following the computations of the angular momentum in GR~\cite{Peters:1964zz, Junker:1992kle, Gopakumar:1997bs, Arun:2009mc}, we derive the angular momentum flux of nonspinning compact binaries up to 1.5PN order in ST theories. 
	These results can be used to compute the radiation-reaction force and hence the waveform for eccentric binaries in ST theories of gravity.
	They can also be implemented in EOB models containing beyond-GR modifications~\cite{Sennett:2016rwa, Julie:2017pkb, Julie:2017ucp, Jain:2022nxs, Jain:2023fvt, Julie:2022qux} in order to perform GR tests using gravitational waves.

	The paper is organized as follows. 
	In Sec.~\ref{sec:STReminder}, we give a brief reminder of ST theories.
	In Sec.~\ref{sec:angmom} we discuss the general structure of angular momentum flux in ST theories. 
	We compute the tensorial contribution to the angular momentum flux including the instantaneous, tail and non-linear memory effects up to 1.5PN order in Sec.~\ref{sec:tensorial} and then the instantaneous and tail contributions due to the scalar in Sec.~\ref{sec:scalar}.
	
	In this work, we use the conventions of Refs.~\cite{Damour:1992we, Damour:1993hw,Damour:1995kt} to define the body-dependent ST parameters (see App.~\ref{app:st_pars} for details).
	Unless other otherwise specified, we will also consider geometric units, such as $G = c = 1$.

	\section{ST theory reminder}
	\label{sec:STReminder}

We consider mono-scalar massless ST theories, in which a scalar field $\varphi$ is minimally coupled, in Einstein frame, to the metric $g_{\mu\nu}$. This theory is described (in the Einstein frame) by the action
\begin{align}
	S&=\frac{c^4}{16 \pi G}\int d^4x \sqrt{-g}(R-2g^{\mu \nu} \partial_\mu \varphi \partial_\nu \varphi)\nonumber\\
	&\qquad\qquad\qquad\qquad+S_m[\Psi, {\mathcal{A}(\varphi)}^2g_{\mu\nu}]\,,
\end{align}
where $\Psi$ collectively denotes the matter fields.
The free dynamics of the Einstein metric $g^{\mu \nu}$ is still determined by the usual Einstein-Hilbert action, while the dynamics of the scalar field originate instead by its coupling to the matter fields. 
This coupling is measured by the parameter 
\begin{equation}
	\label{eq:alpha}
	\alpha \left(\varphi\right) = \frac{d \ln \mathcal{A}\left(\varphi\right)}{d \varphi}\, ,
\end{equation}
where $\mathcal{A}\left(\varphi\right)$ is referred to as ``coupling constant'' and uniquely determines the ST theory.
GR is recovered in the limit $\mathcal{A}\left(\varphi\right) \rightarrow cst$, such that $\alpha \left(\varphi\right) \rightarrow 0$.

In the Jordan frame, the scalar field $\varphi$ couples non-minimally to the Jordan metric $\tilde{g}_{\mu\nu}$, linked to the Einstein metric by a conformal transformation of the type
\begin{equation}
	\label{eq:Jordan_g}
	\tilde{g}_{\mu\nu}={\mathcal{A}(\varphi)}^2 g_{\mu\nu} \, .
\end{equation}
The Jordan frame, where it is a matter that is minimally coupled to $\tilde{g}_{\mu\nu}$, is sometimes called ``physical frame'' 
because the Jordan frame metric determines the dynamics.

When considering compact, self-gravitating objects in ST theories,
we need to take into account that the size and internal gravity of each body depend on the scalar field. 
The scalar field determines the effective gravitational constant.
Following the approach of Ref.~\cite{1975ApJ...196L..59E} (see also \cite{Damour:1992we}), we consider extended bodies as point particles with masses depending on the local value of the scalar field.
In the case of a binary system, with the two bodies labelled as $A$ and $B$, the ``skeletonized'' matter action reads
\begin{equation}
	S_{m}=-c \sum_{I=A,B}\int \sqrt{-g_{\mu \nu}\frac{dx^{\mu}}{d\lambda}\frac{d x^{\nu}}{d\lambda}} m_{I}(\varphi)\,,
\end{equation}
where $m_I\left(\varphi\right)$ is the Einstein-frame masses of each body and $\lambda$ is an affine parameter along the worldlines. Recalling Eq.~\eqref{eq:Jordan_g}, the Jordan-frame masses are defined as
\begin{equation}
	\label{eq:jordan_m}
	\tilde{m}_I \left(\varphi\right) = \frac{m_I\left(\varphi\right)}{\mathcal{A}\left(\varphi\right)}\,.
\end{equation}
In general, since the Jordan-frame (or the Einstein-frame) masses depend on the scalar field $\varphi$, the geodesics will not be universal, thus violating the Strong Equivalence Principle.

The Einstein-frame masses $m_I\left(\varphi\right)$ are used to define dimensionless body-dependent functions parametrizing the scalar field interactions~\cite{Damour:1992we, Damour:1993hw, Damour:1995kt}, namely
\begin{subequations}
	\label{eq:DEF_params}
	\begin{align}
	\alpha_I &=\frac{d\ln m(\varphi)_I}{d\varphi}\,, \\
	\beta_I &=\frac{d\alpha_I}{d\varphi}\,,\\
	\beta'_I &=\frac{d\beta_I}{d\varphi}\,.
\end{align} 
\end{subequations}
In the GR limit of $\mathcal{A}(\varphi)\rightarrow cst.$, all the body-dependent parameters determining the theory vanish, i.e. $\alpha_I = \beta_I = \beta'_I=\beta''_I = 0$.

We will also use index ``0'' to denote quantities evaluated at $\varphi = \varphi_0$, with $\varphi_0$ being the asymptotic value of the scalar field far from the two bodies, e.g. $\alpha_A^0 \equiv \alpha_A \left(\varphi_0\right)$.
We can then define an effective gravitational constant of the binary system as
\begin{equation}
	G_{\rm AB} \equiv \left(1+\alpha^0_A \alpha^0_B\right) G\,.
\end{equation}

In the following, we will use some combinations of the asymptotic ST parameters, such as
\begin{subequations}
\begin{align}
	\delta_{\rm A} &= \frac{\left(\alpha_{\rm A}^0\right)^2}{\left(1+\alpha_{\rm A}^0\alpha_{\rm B}^0\right)^2}\,,  \\
	\bar{\beta}_{\rm A} &= \frac{1}{2}\frac{\left(\alpha_{\rm B}^0\right)^2\beta_{\rm A}^0 }{\left(1+\alpha_{\rm A}^0\alpha_{\rm B}^0\right)^2}\,,  \\
	\mathcal{S}_{\rm A} &= \sqrt{\frac{1+\alpha_0^2}{1+\alpha_{\rm A}^0\alpha_{\rm B}^0}}\left(1-2\alpha_{\rm A}^0\right)\,,  \\
	\theta_{\rm A} &= - \frac{\alpha_{\rm B}^0}{4\alpha_{\rm A}^0} \frac{\beta_{\rm A}^0 \mathcal{S}_{\rm A}}{\left(1+\alpha_{\rm A}^0\alpha_{\rm B}^0\right)}\,, \\
	\Theta_{A} & = \frac{(\alpha_B\beta'_A)^0 \mathcal{S}_B}{8\left(1+\alpha_{A}^0\alpha_{\rm B}^0\right)^2}~,
\end{align}
\end{subequations}
together with the corresponding ones obtained by substituting $A\leftrightarrow B$.
Their (anti-)symmetric combinations are denoted with a subscript (``$-$'')``+'', as in
\begin{equation}
x_{\pm} = \frac{1}{2} \left(x_{\rm A} \pm x_{\rm B}\right)\,.
\end{equation}

Finally, we will make use of the following combinations of ST coefficients:
\begin{subequations}
\begin{align}
	\bar{\gamma}_{\rm AB} &= -2 \frac{\alpha_{\rm A}^0\alpha_{\rm B}^0}{1+\alpha_{\rm A}^0\alpha_{\rm B}^0}\,,  \\
\zeta &= \frac{\alpha_{\rm A}^0 \alpha_{\rm B}^0 \beta_{\rm A}^0 \beta_{\rm B}^0 }{\left(1+\alpha_{\rm A}^0\alpha_{\rm B}^0\right)^3}\,,  \\
\langle \delta \rangle &= \delta_+ + X_{\rm AB} \delta_-\,,  \\
\langle \bar{\beta} \rangle &= \bar{\beta}_+ - X_{\rm AB} \bar{\beta}_-\,,  \\
\langle \theta \rangle &= \theta_+ - X_{\rm AB} \theta_-\,, \\
\langle \Theta \rangle &= \Theta_+ - X_{AB} \Theta_-\,.
\end{align}
\end{subequations}

We report in Appendix.~\ref{app:st_pars} details regarding the relations to Jordan-frame parameters used, e.g., in Refs.~\cite{Bernard:2018hta, Bernard:2018ivi,Bernard:2019yfz, Bernard:2022noq,Bernard:2023eul}.

\section{The far-zone angular momentum flux in ST theories}
	\label{sec:angmom}

In this section, we discuss the basic structure of the angular momentum flux of inspiralling, nonspinning compact binaries in ST theory, 
following the work of Ref.~\cite{Bernard:2022noq}, which computed the energy flux up to 1.5PN order.
We here define the $n$-th PN order (following Ref.~\cite{Bernard:2022noq}) as the term proportional to $1/c^{2n}$ beyond the leading GR quadrupole term. 
This means the leading-order ST dipolar term will be denoted as -1PN order.

The angular momentum flux in ST theories can be expressed as the sum of a tensorial and a scalar contribution, such that 
\begin{equation}
	\mathcal{G}_i = \mathcal{G}_i^{\rm T} + \mathcal{G}_i^{\rm S} \equiv \left(\frac{{\rm d} L_i}{{\rm d}  t}\right)^{\rm GW}~,
\end{equation}
where the superscripts ``S" and ``T" denote the scalar and tensorial contributions respectively.

The explicit expressions of angular momentum flux are derived using the radiative multipole moments. 
We will compute the tensorial angular momentum flux $\mathcal{G}_i^{\rm T}$ up to 
1.5PN order, including both instantaneous and hereditary
contributions (entering at 1.5PN order).
Instead, we explicitly compute the scalar flux $\mathcal{G}_i^{\rm S}$ only up to 1PN as for the instantaneous contribution at 1.5PN order one must need the explicit expression of scalar dipole moments at relative 2.5PN accuracy which are not explicitly given in Ref.~\cite{Bernard:2022noq}. 

We describe two particles with relative separation $x^i$ and relative velocity $v^i$. 
We will use dots to denote time derivatives, e.g. $v^i=\dot{x}^i\equiv{\rm d}x^i/{\rm d}t\,$.
The angular momentum of the system is then computed as $L_i = \epsilon_{i j k} x^j v^k$.
When working with polar coordinates $(r,\varphi')$, we define the unit direction along the angular momentum as 
\begin{equation}
	\hat{L}_i \equiv \frac{\epsilon_{i j k} x^j v^k}{r^2 \dot{\varphi'} }\,,
\end{equation}
and the total velocity of the system as $v = \sqrt{\dot{r}^2 + r^2\dot{\varphi'}^2}$.

The angular momentum flux $\mathcal{G}_i$ is orthogonal to the orbital plane, and aligned with the angular momentum.  
We can then compute the modulus of the angular momentum flux $\mathcal{G}$, such that 
\begin{equation}
	\mathcal{G}_i = \mathcal{G} \hat{L}_i\,.
\end{equation}

\section{Tensorial contribution}
	\label{sec:tensorial}

	The tensorial contribution to the angular momentum flux up to 1.5PN is expressed through the formulas derived in Ref.~\cite{Thorne:1980ru}, namely
 \begin{align}
 \label{eq:tens-AngMom}
\mathcal{G}_i^{\rm T} &= \frac{G\phi_0}{c^5} \epsilon_{i j k} \Bigg\{\frac{2}{5}\mathcal{U}_{j \ell} \,\mathcal{\dot{U}}_{k \ell} \nonumber \\
&+ \frac{1}{c^2}\left[\frac{1}{63}\mathcal{U}_{j \ell m}\,\mathcal{\dot{U}}_{k \ell m} + \frac{32}{45}\mathcal{V}_{j \ell}\,\mathcal{\dot{V}}_{k \ell}\right] + O\left(\frac{1}{c^4}\right)\Bigg\}\,,
 \end{align}
where $\mathcal{U}$ and $\mathcal{V}$ are, respectively, the symmetric-trace-free (STF) mass and current radiative, tensorial, multiple moments.
Only the mass quadrupole multipole moment, $\mathcal{U}_{j \ell} $, differs by its twice differentiated source counterpart up to 1.5PN order (see, Refs.~\cite{Lang:2013fna, Lang:2014osa, Bernard:2022noq}), i.e.,
\begin{align}
\label{eq:rad-source-tens}
\mathcal{U}_{j \ell}  &= \ddot{I}_{j \ell}+ \frac{2 GM}{\phi_0c^3}\int^{T_{\rm R}}_{-\infty}dV \overset{(4)}{I_{j \ell}}(V)\left[\ln\left(\frac{T_{\rm R}-V}{2\tau_0}\right)+\frac{11}{12}\right] \nonumber \\
&+\frac{G(3+2\omega_0)}{3c^3}\left(\int^{T_{\rm R}}_{-\infty}dV\ddot{I}^{\rm S}_{\langle j}(V)\ddot{I}^{\rm S}_{\ell \rangle}(V)-A_{j\ell}\right)\nonumber \\
&+ O\left(\frac{1}{c^4}\right)\,,
\end{align}
where 
\begin{equation}
A_{j \ell} = {I^{\rm S}_{\langle j}}\overset{(3)}{I^{\rm S}_{\ell \rangle}}+\dot{I}^{\rm S}_{\langle j}\ddot{I}^{\rm S}_{\ell \rangle}+\frac{1}{2}{I^{\rm S}}\overset{(3)}{I^{\rm S}_{j\ell }}\,,
\end{equation}
and we introduced the superscripts ``($n$)" to denote $n$-th time derivatives.
Hereafter, the angular brackets denote the STF part.  The quadrupole radiative moment $\mathcal{U}_{ij}$ given in Eq.~\eqref{eq:rad-source-tens} is a combination of two contributions: (i) instantaneous contribution, which is a function of retarded time $T_{\rm R}$, and (ii) hereditary contribution, which depends on the dynamics of the system in the entire past, i.e. $V\leq T_{\rm R}$. The parameter $\tau_0$ is a constant time scale that enters the relation between the retarded time $T_{\rm R}$ in radiative coordinates and the corresponding time in harmonic coordinates.
Therefore, from the above equation, we observe that starting at 1.5PN, the tensorial part of the angular momentum flux also contains both instantaneous and hereditary (tail and memory) contributions, so that we can write
\begin{equation}
	\mathcal{G}^{\rm T} = \mathcal{G}^{\rm inst, T} + \mathcal{G}^{\rm hered, T}.
\end{equation}

The hereditary contributions can then also be further decomposed as,
	\begin{equation}
	\mathcal{G}_i^{\rm hered, T} = \mathcal{G}_i^{\rm  tail, T}+\mathcal{G}_i^{\rm mem, T}~,
	\end{equation}
	where $\mathcal{G}_i^{\rm mem, T}$ is the non-linear memory effect and $\mathcal{G}_i^{\rm tail, T}$ is the tail effect.
	We will derive the explicit expressions of these contributions in the following subsections.

\subsection{Instantaneous contribution}

The instantaneous tensorial term entering the angular momentum flux, up to 1.5PN in terms of the source multipole moments following Eq.~\eqref{eq:tens-AngMom} is,

\begin{align}
\label{eq:tensMM}
\mathcal{G}^{\rm inst, T}  =&\frac{G\phi_0}{c^5}\epsilon_{ijk}\left\{\frac{2}{5} \ddot{I}_{ja}\overset{(3)}{I_{ka}}+\frac{1}{c^2}\left(\frac{1}{63}\overset{(3)}{I_{jab}}\overset{(4)}{I_{kab}}+\frac{32}{45}\ddot{J}_{ja} \overset{(3)}{J_{ka}}\right)
\right.\nonumber\\
&\left.-\frac{2G(3+2\omega_0)}{15c^3}\left(A_{ja}\overset{(3)}{I_{ka}}+\dot{A}_{ka}\ddot{I}_{ja}\right)\right.\nonumber\\
&\left.+\frac{2G(3+2\omega_0)}{15c^3}\ddot{I}_{ja} \ddot{I}^{\rm S}_{\langle k} \ddot{I}^{\rm S}_{a\rangle}\right\}~.
\end{align}
Here, $I_{ij}$ and $I_{ijk}$ are the tensorial source quadrupole and octupole multipole moments and $I^S$, $I^S_i$ and $I^S_{ij}$ are the scalar source monopole, dipole and quadrupole multipole moments.

This instantaneous tensorial term entering the angular momentum flux up to 1.5PN can be written as,
	\begin{equation}
		\mathcal{G}^{\rm inst, T}= \mathcal{G}^{\rm inst,T}_{\rm N} +\mathcal{G}^{\rm inst, T}_{\rm 0.5PN} +\mathcal{G}^{\rm inst,T}_{\rm 1PN}+\mathcal{G}^{\rm inst,T}_{\rm 1.5PN}~.
	\end{equation}
	The contributions up to 1.5PN order in Harmonic coordinates are found to be
	\begin{widetext}
		\begin{subequations}
		\begin{align}
			\label{eq:GT}
			\mathcal{G}^{\rm inst,T}_{\rm N} &=\frac{G_{\text{AB}}^2M^3 \nu ^2 }{c^5r} \dot{\varphi'}  \left(1+\frac{\gamma_{\rm AB}}{2}\right)\frac{8}{5} \left(2\frac{G_{\text{AB}} M}{r} + 2 { v^2}-3{ \dot{r}^2}\right)~,\\
			\mathcal{G}^{\rm inst,T}_{\rm 0.5PN} &= 0~, \\
			\mathcal{G}^{\rm inst,T}_{\rm 1PN} &= \frac{G_{\rm AB}^2 M^3 \nu^2}{c^7r} \dot{\varphi'}\left(1+\frac{\gamma_{\rm AB}}{2}\right)\frac{8}{5}
			\left\{\left(-\frac{745}{42}+\frac{\nu}{21}-7\gammaAB-8\langle\bar{\beta}\rangle\right)\left(\frac{G_{\rm AB}M}{r}\right)^2+
			\left(-\frac{58}{21} -\frac{95}{21} \nu-4\langle\bar{\beta}\rangle\right)\frac{G_{\rm AB}M}{r} v^2+
			\right.\nonumber\\
			&\left.+ \left(\frac{62}{7}+\frac{197}{42}\nu+2\gammaAB+8\langle\bar{\beta}\rangle\right)\frac{G_{\rm AB}M}{r} \dot{r}^2+\left(\frac{307}{84}-\frac{137}{21}\nu+\gamma \right)v^4
			+\left(-\frac{37}{7}+\frac{277 }{14}\nu\right)v^2\dot{r}^2+\left(\frac{95}{28} - \frac{90}{7}\nu\right)\dot{r}^4\right\}\,,\\
			\label{eq:tens-inst15pn}
			\mathcal{G}^{\rm inst,T}_{\rm 1.5PN} &=\frac{G_{\rm AB}^2M^3\nu^2}{c^8r}\dot{\varphi'}\left(1+\frac{\gammaAB}{2}\right)\left(\frac{8}{5}\zeta\, \frac{G_{\rm AB}M}{r} \dot{r}\right)\left\{ \left[\mathcal{S}_+^2-\left(1-\frac{50}{3}\nu\right)\mathcal{S}_-^2\right]\frac{G_{\rm AB}M}{r}+2\left[\mathcal{S}_+^2-\left(1-14\nu\right)\mathcal{S}_-^2\right]v^2\right.\nonumber\\
			&\left.-\frac{5}{2}\left[\mathcal{S}_+^2-\left(1-16\nu\right)\mathcal{S}_-^2\right]\dot{r}^2\right\}\,.
		\end{align}
	\end{subequations}
	\end{widetext}

The above results also contain an instantaneous contribution coming from the memory term of Eq.~\eqref{eq:rad-source-tens} entering the angular momentum flux. The memory term of Eq.~\eqref{eq:rad-source-tens} is just an anti-derivative term, therefore, one of its contributions corresponding to the time derivative of $\mathcal{{U}}_{k l}$ in Eq.~\eqref{eq:tens-AngMom} becomes instantaneous.  This contribution enters at 1.5PN order and hence, is incorporated in the 1.5PN instantaneous contribution given above [see  Eq.~\eqref{eq:tens-inst15pn}]. There is also a remaining memory effect entering at 1.5PN order which we will discuss in Sec.~\ref{sec:MemE}.
On the contrary, there are no memory contributions in the energy flux as in the energy flux formula the memory is completely time-differentiated 
and therefore becomes instantaneous (see, Ref.~\cite{Bernard:2022noq}).

The results obtained in this section match with the angular momentum flux computed in Refs.~\cite{Peters:1964zz, Junker:1992kle,Gopakumar:1997bs, Arun:2009mc} in the GR limit, i.e. $\alpha_{A,B}^0\rightarrow0$.

	\subsection{Hereditary Contributions}
	\label{sec:tensorial-nonloc}
Now, we move on to compute the hereditary contribution to the tensorial part of angular momentum flux. As for the flux computations in GR~\cite{Gopakumar:1997bs, Arun:2009mc}, it is not possible to obtain the closed-form results for these contributions in the time domain, therefore we adopt the same strategy used in GR of using the discrete Fourier decomposition in ST theory.

As mentioned previously, the hereditary contribution can be further decomposed as,
\begin{equation}
	\mathcal{G}_i^{\rm hered,T} = \mathcal{G}_i^{\rm tail, T}+\mathcal{G}_i^{\rm mem, T}
\end{equation}

The tail and memory integrals entering in the tensorial angular momentum up to 1.5PN order in ST theory are given by
\begin{align}
\label{eq:tens-tail}
\mathcal{G}_i^{\rm tail, T} &= \frac{4 G^2 M}{5c^8}\epsilon_{ijk}
\nonumber \\
&\times\left\{\ddot{I}_{j \ell}\int_{-\infty}^{T_{\rm R}}dV \overset{(5)}{I_{k \ell}}(V)\left[\ln\left(\frac{T_{\rm R}-V}{2\tau_0}\right)+\frac{11}{12}\right]
\right.\nonumber \\
&\left.-\overset{(3)}{I_{j \ell}}\int_{-\infty}^{T_{\rm R}}dV \overset{(4)}{I_{k \ell}}(V)\left[\ln\left(\frac{T_{\rm R}-V}{2\tau_0}\right)+\frac{11}{12}\right]\right\}\,,
\end{align}
and,
\begin{align}
\label{eq:tens-mem}
\mathcal{G}_i^{\rm mem, T} & = \frac{2 G^2 (1-\zeta')}{15\zeta' c^8}\epsilon_{ijk}
{I}^{(3)}_{j\ell}\int_{-\infty}^{T_{\rm R}}dV \ddot{I}^{\rm \,S}_{\langle{k}}(V){\ddot{I}^{\rm \,S}}_{{\ell}\rangle}(V)\,.
\end{align}
Here the angular brackets denote the STF part. 

From Eq.~\eqref{eq:tens-tail}, we observe that the leading order tail contribution entering at 1.5PN order is the usual GR tail contribution \cite{Gopakumar:1997bs, Rieth:1997mk, Arun:2009mc, Racine:2008kj} and the corrections due to the scalar-field only enters as a memory effect at 1.5PN order (see, Eq.~\eqref{eq:tens-mem} for its explicit expression). Therefore, in this section, we will focus only on computing the memory effect contribution.


\subsubsection{Memory contribution}
\label{sec:MemE}

The memory contribution to the tensorial part of angular momentum flux from Eq.~\eqref{eq:tens-mem} is given by;
\begin{align}
\label{eq:tens-mem1}
\mathcal{G}_i^{\rm mem, T} & = \frac{2 G^2 (1-\zeta')\phi_0}{15\zeta' c^8}\epsilon_{ijk}
\overset{(3)}{I_{j\ell}}(\mathrm{T_{R}})\int_{-\infty}^{T_{\rm R}} dV \ddot{I}^{\rm \,S}_{{k}}(V){\ddot{I}^{\rm \,S}}_{{\ell}}(V)\,.
\end{align}
The above equation is symmetric with respect to the indices $k$ and $\ell$ and is also trace-free due to the presence of $\epsilon_{ijk}$, hence we have omitted the angular brackets from the indices $k$ and $\ell$ used to indicate the STF projection.  The orbital average therefore reads,
\begin{align}
\langle\mathcal{G}_i^{\rm mem, T} \rangle & = \frac{2 G^2 (1-\zeta')\phi_0}{15\zeta'c^8}\epsilon_{ijk}  \nonumber \\
&\times
\int_0^P\frac{dT_{\rm R}}{P}\overset{(3)}{I_{j\ell}}(\mathrm{T_{R}})\int_{-\infty}^{T_{\rm R}}dV \ddot{I}^{\rm \,S}_{{k}}(V){\ddot{I}^{\rm \,S}}_{{\ell}}(V)\,,
\end{align}
which can be re-expressed as,
\begin{align}
\label{eq:reexp}
\langle\mathcal{G}_i^{\rm mem, T} \rangle & = \frac{2 G^2 (1-\zeta')\phi_0}{15\zeta'c^8}\epsilon_{ijk}  \nonumber \\
&
\left\{\int_{-\infty}^{0}dV \ddot{I}^{\rm \,S}_{{k}}(V){\ddot{I}^{\rm \,S}}_{{\ell}}(V)\int_0^P\frac{dT_{\rm R}}{P}\overset{(3)}{I_{j\ell}}(\mathrm{T_{R}}) \right.\nonumber \\
&\left.+ \int_{0}^{P}dV \ddot{I}^{\rm \,S}_{{k}}(V){\ddot{I}^{\rm \,S}}_{{\ell}}(V)\int_V^P\frac{dT_{\rm R}}{P}\overset{(3)}{I_{j\ell}}(\mathrm{T_{R}})\right\}\,
\end{align}

In the first instance, the contribution due to the first term appears to be zero as this term contains the orbital average of the third time-derivative of the Newtonian tensorial quadrupole moment up to 1.5PN order and as the motion is periodic, the quadrupole averages to zero. However, as mentioned in Ref.~\cite{Arun:2009mc}, we can not replace this term with zero because this argument neglects the evolution of the Keplerian orbital elements in the remote past. We study the impact of this term and compute the non-zero DC contribution to the memory effects in ST theory. 

Whereas, the contribution of the second term, i.e., due to the recent past of the source averages to zero. This implies that the non-zero DC contribution is due to the remote past of the source, i.e., the first term of Eq.~\eqref{eq:reexp}.

Now, we compute the DC contribution of the non-linear memory integral following the method of Ref.~\cite{Arun:2009mc}. For the leading order contribution to the memory integral, i.e., 1.5PN order, it is sufficient to replace the tensorial quadrupolar moment and scalar dipole moment and their time derivatives with their Newtonian values. Here, we work in polar coordinates $\left(r,\varphi'\right)$ and express Eq.~\eqref{eq:tens-mem1} in terms of the current orbital separation $r_0$, radial velocity $\dot{r}_0$, orbital frequency $\dot{\varphi'}_0$ and orbital phase $\varphi'_0 =0$. For any instant in the past, we will use the notation $r, ~\dot{r}~\textrm{and}~\dot{\varphi'}$ to denote the orbital separation, radial velocity and orbital frequency, respectively. From this, we obtain
\begin{align}
\label{eq:memDC}
\mathcal{G}^{\rm mem, T} &= -\frac{8\nu^3}{15 r_0}\frac{G_{AB}^5M^6\mathcal{S}_-^2(1-\zeta')\zeta'}{\alpha_{AB}}
\nonumber\\
&\times\int_{t_1}^{T_{\rm R}}dV\left(4\frac{\dot{\varphi'}_0}{r^4}\cos(2\varphi')-\frac{\dot{r}_0}{r_0r^4}\sin(2\varphi')\right)\,~,
\end{align}
where we denote by $t_1$ some initial instant of the formation of the system with some eccentricity close to 1.

We then replace $r$, $\dot{r}$ and $\dot{\varphi'}$ in the above equation by the explicit solution for Keplerian motion in ST theories, i.e.,
\begin{align}
r &= \frac{a(1-e^2)}{1 + e \cos\varphi'}~,\\
\dot{r} &= \sqrt{\frac{G_{ AB}}{a(1-e^2)}}e \sin \varphi'~,\\
\dot{\varphi'} &= \sqrt{\frac{G_{ AB}}{a^3(1-e^2)^3}}(1+e \cos\varphi')^2~.
\end{align}
On inserting the above equations in Eq.~\eqref{eq:memDC}, we obtain an integrand as a function of semi-major axis $a(t)$, eccentricity $e(t)$ and orbital phase $\varphi(t)$ such that the resulting integral is composed of many harmonics of the Keplerian motion in terms of orbital phase $\varphi(t)$. The orbital phase oscillates at the rate of the orbital period in comparison to $a(t)$ and $e(t)$ which remain approximately constant during one period but
slowly evolved by radiation reaction during the past evolution of the binary. We expect that the
oscillations of the orbital phase, due to the sequence of many orbital cycles in the entire life of the
the system will more or less cancel each other in the memory integral so that the true PN
order of the oscillating terms will be simply given by the power of $1/c$ they carry. 

The explicit expression then becomes
\begin{align}
\mathcal{G}^{\rm mem, T} = -\frac{8\nu^3\dot{\varphi'}_0}{15r_0}\frac{G_{AB}^5M^6\mathcal{S}_-^2(1-\zeta')\zeta'}{\alpha_{AB}}\int_{t_1}^{t_0}dt\frac{e^2(e^2+6)}{a^4(e^2-1)^4}\,.
\end{align}

\section{Scalar Contribution}
	\label{sec:scalar}
	In order to derive the scalar contribution to the angular momentum flux in massless ST theories, we need to find its relation to radiative scalar multipole moments. The corresponding formula, linking the scalar contribution to the \textit{energy} flux in terms of radiative multipole moments is given in Eq.~(3.27b) of Ref.~\cite{Bernard:2022noq}. 
	
	The angular momentum losses can be easily calculated using the energy-momentum tensor for scalar waveform. 
	The scalar waveform for the massless ST theory is given by Eq.~(3.25) of Ref.~\cite{Bernard:2022noq} and reads
	\begin{equation}
		\psi = -\frac{2 G}{c^2 R} \sum_{\ell=0}^{\infty}\frac{1}{c^\ell \ell!} N_{i_1 i_2...i_\ell}~\mathcal{U}^{\rm S}_{i_1 i_2...i_\ell}\,,
	\end{equation}
	where $\mathcal{U}^{\rm S}_{i_1 i_2...i_\ell}$ is the scalar radiative moment, and $N_{i_1 i_2...i_\ell} \equiv N_{i_1}...N_{i_\ell}$. $N_i$ is a versor of the polarization orthonormal triad $(\mathbf{N}, \mathbf{P}, \mathbf{Q})$~\cite{Faye:2012we,Bernard:2022noq}. 
	The energy-momentum tensor for the scalar GWs corresponding to the action of Ref.~\cite{Bernard:2022noq} is,
	\begin{equation}
		T_{\alpha \beta}^{\rm S, GW}= \frac{c^3 \phi_0 (1-\zeta')}{16\zeta' \pi G}\langle \partial_{\alpha}\psi \partial_{\beta} \psi\rangle~.
	\end{equation}
	Hence,  the scalar angular momentum flux in ST theory is (see, Ref.~\cite{Thorne:1980ru} for GR)
	\begin{align}
		\label{eq:sangmom1}
		\mathcal{G}^{\rm S}_i &= - \int R^2 \epsilon_{ijk}x_jT^{\rm S, GW}_{0k} d\Omega~,\nonumber \\
		&=-\frac{c^3 R^2 \phi_0 (1-\zeta')}{16\zeta' \pi G}\int \epsilon_{ijk} x_j \partial_0 \psi \partial_k \psi d\Omega~.
	\end{align} 
	Now using,
	\begin{align}
		\partial_0 \psi &= -\frac{2G}{R c^2} \sum_{\ell=0}^{\infty} \frac{1}{c^\ell \ell!} N_{i_1 i_2...i_\ell} \mathcal{\dot{U}}^{\rm S}_{i_1 i_2...i_\ell}\\
		\partial_k \psi &= -\frac{2G}{R^2 c^2} \sum_{\ell=0}^{\infty} \frac{\ell+1}{c^{\ell+1} (\ell+1)!} N_{i_1 i_2...i_\ell}~\mathcal{U}^{\rm S}_{k i_1 i_2...i_\ell} \nonumber \\
		&+\frac{2G}{R^2 c^2} \sum_{\ell=0}^{\infty} \frac{ n_k}{c^{\ell+1} (\ell+1)!} N_{i_1 i_2...i_{\ell+1}}~\mathcal{U}^{\rm S}_{ i_1 i_2...i_{\ell+1}}
	\end{align}
	Substituting these values in Eq.~\eqref{eq:sangmom1}, we obtain
	\begin{widetext}
		\begin{equation}
			\mathcal{G}^{\rm S}_i = -\frac{ G \phi_0 (1-\zeta')}{4 \zeta' \pi c} \epsilon_{ijk}n_j\int d\Omega \sum_{\ell=0}^{\infty}\sum_{\ell'=0}^{\infty} \frac{1}{c^\ell c^{\ell'+1} \ell! (\ell'+1)!} 
			\left[(\ell'+1)N_{A_l}N_{B_{\ell'}} \mathcal{\dot{U}}^{\rm S}_{A_\ell} \mathcal{\dot{U}}^{\rm S}_{k B_{\ell'}}-n_k N_{A_{\ell}}N_{B_{\ell'+1}} \mathcal{\dot{U}}^{\rm S}_{A_\ell} \mathcal{\dot{U}}^{\rm S}_{B_{\ell'+1}}\right]\,.
		\end{equation}
	\end{widetext}
	Integrating the above equation using Eq.~(2.6) of Ref.~\cite{Thorne:1980ru},  we find that the only non-zero contribution to the above equation is from the first term in the bracket when $l'=l-1$, i.e.
	\begin{align}
	\label{eq:AngMs}
		\mathcal{G}^{\rm S}_i = \epsilon_{ijk}\sum_{l=1}^{\infty}\frac{G\phi_0(1-\zeta')}{\zeta' c^{2l+1}(l-1)! (2l+1)!!}\dot{\mathcal{U}}^{\rm S}_{j i_1 i_2...i_{l-1}} \mathcal{U}^{\rm S}_{k i_1 i_2...i_{l-1}} \,.
	\end{align}
	
	Substituting the expression of the scalar radiative moment from Ref.~\cite{Bernard:2022noq} in the above equation, we obtain the scalar contribution to the angular momentum flux. Note that, similar to the tensorial contribution to the angular momentum flux, the scalar contribution to the angular momentum flux also contains both instantaneous and tail contributions, so that we can write
\begin{equation}
	\mathcal{G}^{\rm S} = \mathcal{G}^{\rm inst, S} + \mathcal{G}^{\rm tail, S}~,
\end{equation}
where $\mathcal{G}^{\rm inst, S}$ and $ \mathcal{G}^{\rm tail, S}$ are the instantaneous and tail contribution to the scalar angular momentum flux, respectively.
	This is because the monopole scalar radiative moment $\mathcal{U}_{j}^{\rm S}$ entering Eq.~\eqref{eq:AngMs} differs by its time differentiated source part ($I^{\rm S}_i$) up to 1.5PN order,
	\begin{align}
	\mathcal{U}^{\rm S}_i = \dot{ I^{\rm S}_i} + \frac{2 G M}{\phi_0 c^3}\int_{-\infty}^{T_{\rm R}}dV\overset{(4)}{I_k^{\rm S}}(V)\left[\ln\left(\frac{T_{\rm R}-V}{2\tau_0}\right)+1\right]~.
\end{align}	
	\subsection{Instantaneous contribution}
	\label{sec:GSinst}
	
	The instantaneous scalar contribution at 1.5PN order is,
	\begin{equation}
		\label{eq:G}
		\mathcal{G}^{\rm inst, S}= \mathcal{G}^{\rm inst,S}_{\rm -1PN} +\mathcal{G}^{\rm inst,S}_{\rm N} +\mathcal{G}^{\rm inst, S}_{\rm 0.5PN} +\mathcal{G}^{\rm inst,S}_{\rm 1PN}+\mathcal{G}^{\rm inst,S}_{\rm 1.5PN}
	\end{equation}
	where the Newtonian contribution corresponds to the dominant contribution in GR at 2.5PN order.
	The results at 1PN order in Harmonic coordinates are
	\begin{widetext}
		\begin{align}
			\label{eq:G-1PN}
			\mathcal{G}^{\rm inst,S}_{\rm -1PN} &= \frac{G_{AB}^2 M^3 \nu^2}{c^3 r}\dot{\varphi'}\left(\frac{4}{3}\zeta' \mathcal{S}_-^2\right)\,,\\
			\mathcal{G}^{\rm inst,S}_{\rm N}& =\frac{G_{AB}^2M^3\nu^2}{c^5r}\dot{\varphi'}~\left(\frac{2}{3}\zeta'\right)\left\{\left[\frac{1}{5}\left(\left(-53-20\gammaAB + 3\nu\right)\Sm^2 +4\Sp^2-2\Sm \Sp\left(3X_{AB}-10\bar{\beta}_-+10X_{AB}\bar{\beta}_+ \right)\right)\right.\right.
			\nonumber\\
			&\left.\left.-4(-1+\gammaAB)\langle\theta\rangle \Sm \right]\frac{G_{AB}M}{r}+\left[\left(\frac{9}{5}+\frac{24}{5}\nu\right)\mathcal{S}_-^2-\frac{18}{5}X_{AB}\mathcal{S}_+ \mathcal{S}_- -12\langle \theta\rangle \mathcal{S}_- \frac{6}{5}\mathcal{S}_+^2\right]\dot{r}^2
			\right.
			\nonumber\\
			&\left.+ \left[\frac{6}{5}X_{AB}\mathcal{S}_+\mathcal{S}_- +\left(3+\frac{3}{5}\nu+2\gammaAB-\frac{13}{5}\nu \right)\mathcal{S}_-^2 +4\langle \theta\rangle \mathcal{S}_-+\frac{4}{5}\mathcal{S}_+^2\right]v^2\right\}\,,\\
			\mathcal{G}^{\rm inst,S}_{\rm 0.5PN}& =\frac{G_{AB}^2M^3\nu^2}{c^6r}\dot{\varphi'}~\left(\frac{2\zeta' \mathcal{S}_-^2}{3}\right) \left\{\gammaAB -4 \zeta' \left[X_{AB}    \mathcal{S}_+ - (4 \nu -1) \mathcal{S}_-\right]\mathcal{S}_-\right\} \frac{G_{AB}M}{r}\dot{r}\,,\\
\label{eq:GS}
			\mathcal{G}^{\rm inst,S}_{\rm 1PN}& =\frac{G_{AB}^2M^3\nu^2}{c^7r}\dot{\varphi'}\left(\zeta'\right)
			\left\{\mathcal{G}^{\rm r^2}_{\rm 1PN}\left(\frac{G_{\rm AB }M}{r}\right)^2+\mathcal{G}^{\rm r,v^2}_{\rm 1PN}\left(\frac{G_{\rm AB }M}{r}\right)v^2+\mathcal{G}^{\rm r,\dot{r}^2}_{\rm 1PN}\left(\frac{G_{\rm AB }M}{r}\right)\dot{r}^2+\mathcal{G}^{\rm v^4}_{\rm 1PN}v^4\right.
			\nonumber\\
			&\left.+\mathcal{G}^{\rm v^2\dot{r}^2}_{\rm 1PN}v^2\dot{r}^2+\mathcal{G}^{\rm \dot{r}^4}_{\rm 1PN}\dot{r}^4\right\}\,,
		\end{align}
	\end{widetext}
where the 1PN coefficients explicitly read
	\begin{widetext}
	\begin{align}
	\mathcal{G}^{\rm r^2}_{\rm 1PN} =&\frac{2}{525}\left(-1171+5134\nu-490\gammaAB+1960\gammaAB\nu-1464\nu^2-350X_{AB}(9+4\gammaAB-6\nu)\bar{\beta}_-\right.
	\nonumber\\
	&\left.+350X_{AB}\delta_-\right)\Sm^2+\frac{2}{525}\left(280X_{AB}(10+\gammaAB(2-8\nu)+\nu)\thm+140(-3+4\gammaAB)X_{AB}^2\thp\right)\Sm
	\nonumber\\
	&+\frac{4}{105}\left(28(-5+8\nu)\bar{\beta}_-+X_{AB}(212+77\gammaAB-39\nu+28(5+8\nu)\bar{\beta}_+)\right)\Sp
	\nonumber\\
	&+\frac{2}{105}\left(-251-98\gammaAB+18\nu\right)\Sp^2
	+\frac{32}{3}\gammaAB X_{AB}\thm\thp+\frac{8}{15\left(\Sm^2-\Sp^2\right)^2}\left[-2X_{AB}\thm\Sm^5\right.
		\nonumber\\
	&\left.+2\thp\left(X_{AB}\Sp+20\gammaAB(1-2\nu)\thp\right)\Sp^4+\left(10\gammaAB(-1+10\nu)\thm^2-21\thm\Sp+2X_{AB}\Sp\thp\right)\Sm^4\right.
	\nonumber\\
	&\left.+\left(10\gammaAB X_{AB}\thm^2+\thp(21\Sp-10\gammaAB X_{AB}\thp)+4\thm(X_{AB}\Sp+5\gammaAB(1-10\nu)\thp)\right)\Sm^3\Sp\right.
	\nonumber\\
	&\left.-\left(10\gammaAB X_{AB}\thm^2+\thp(21\Sp-10\gammaAB X_{AB}\thp)+2\thm(X_{AB}\Sp+20\gammaAB(1-2\nu)\thp)\right)\Sm\Sp^3\right.
	\nonumber\\
	&\left.\left(20\gammaAB(1-2\nu)\thm^2+21\thm\Sp+2\thp(-2 X_{AB}\Sp+5\gammaAB(-1+10\nu)\thp)\right)\Sm^2\Sp^2\right]
	\nonumber\\
	&+\frac{\alpha_{AB}^2}{(1-\zeta')^2}\left\{\frac{\Sm}{8400}\left[\Sm\left(169392+7892(2+\gammaAB)^2\nu^2-16\left(-18707-7630\bar{\beta}_++700\bar{\beta}_+^2\right.\right.\right.\right.
	\nonumber\\
	&\left.\left.\left.\left.-700\delta_+-350\langle\epsilon\rangle\right)\gammaAB+4\left(49367+17500\bar{\beta}_++700\delta_++350\langle\epsilon\rangle\right)\gammaAB^2+560\left(103+20\bar{\beta}_+\right)\gammaAB^3\right.\right.\right.
	\nonumber\\
	&\left.\left.\left.+6300\gammaAB^4+6720\bar{\beta}_+-44800\bar{\beta}_+^2+11200\delta_++5600\langle\epsilon\rangle+\left(-63808+5600(2+\gammaAB)^2\bar{\beta}_+\right.\right.\right.\right.
	\nonumber\\
	&\left.\left.\left.\left.-110848\gammaAB-79092\gammaAB^2-27860\gammaAB^3-4025\gammaAB^4-67200(4+\gammaAB)\bar{\beta}_+^2-42000(\delta_-^2-\delta_+^2)\right.\right.\right.\right.
	\nonumber\\
	&\left.\left.\left.\left.-2800(2+\gammaAB)^2(2\delta_++\epsilon_+)\right)\nu\right)-94080\thp\right]-\frac{8\Sm}{15\left(\Sm^2-\Sp^2\right)^2}\left[-21\Sp^4\thp\right.\right.
	\nonumber\\
	&\left.\left.+\thm\left(10\gammaAB(1+6\nu)\thm+21\Sp\right)\Sm\Sp^2+\thp\left(-20\gammaAB(1+6\nu)\thm+21\Sp\right)\Sm^2\Sp\right.\right.
	\nonumber\\
	&\left.\left.+\left(-21\thm\Sp+10\gammaAB(1+6\nu)\thp^2\right)\Sm^3\right]\right\}~,
	\nonumber\\
		\mathcal{G}^{\rm r, \dot{r}^2}_{\rm 1PN} =&\frac{32}{3}X_{AB}\thm\thp-\frac{2}{105}\left[\left(-170+897\nu-428\nu^2+\gammaAB(-28+112\nu)+X_{AB}(56(-4+3\nu)\bar{\beta}_-\right.\right.
		\nonumber\\
		&\left.\left.+140\delta_-)\right)\Sm^2+\left(28(-13+4\gammaAB)X_{AB}^2\thav + 14 X_{AB}(155-213)\thm\right)\Sm+\left(-112(-7+24\nu)\bar{\beta}_-\right.\right.
		\nonumber\\
		&\left.\left.+X_{AB}(249+804\nu+413\gammaAB+112(-7+4\nu)\bar{\beta}_+)\right)\Sm\Sp+\left(-184+89\nu-28\gammaAB\right)\Sp^2\right.
		\nonumber\\
		&\left.\left(28(-5+4\gammaAB)X_{AB}\thav-1120X_{AB}\Thav+112\thm+280\nu\thm-448\nu\gammaAB\thm+2240\nu\Thm\right)\Sp\right]
		\nonumber\\
		&+\frac{1}{3}\left[\frac{64(-1+2\nu)\Sp\thp\left(\Sm\thm-\Sp\thp\right)}{\Sm^2-\Sp^2}\right]-\frac{4}{15\left(\Sm^2-\Sp^2\right)^2}\left[8X_{AB}\thm\Sm^5\right.
		\nonumber\\
		&\left.+\left((20-720\nu)\thm^2-8X_{AB}\thav\Sp+19(-9+7\nu)\thm\Sp\right)\Sm^4+\left(-80(-1+2\nu)X_{AB}\thm^2\right.\right.
		\nonumber\\
		&\left.\left. +X_{AB}(-155+101\nu)\thm\Sp-80((-1+2\nu)\Thav+2X_{AB}\nu \Thm)\Sp+\thav(-80(-1+2\nu)\thm\right.\right.
		\nonumber\\
		&\left.\left.+(-163+101\nu)\Sp)\right)\Sm\Sp^3+\left(20(-3+8\nu)\thm^2+(163-101\nu)\thm\Sp+80(-12\nu)\Thm\Sp\right.\right.
		\nonumber\\
		&\left.\left.+16X_{AB}\Sp\thp+40\thp^2-800\nu\thp^2-80X_{AB}\Sp\Thp\right)\Sp^2\Sm^2+4\left((5-20\nu)\thm^2\right.\right.
		\nonumber\\
		&\left.\left.+20(-1+2\nu)\Thm\Sp-2X_{AB}\Sp\thp-15\thp^2+40\nu\thp^2+20X_{AB}\Sp\Thp\right)\Sp^4\right.
		\nonumber\\
		&\left.+\left(40(-1+36\nu)\thav\thm+40X_{AB}(-1+36\nu)\thm^2\-16X_{AB}\thm\Sp+80\Thm X_{AB}\Sp \right.\right.
		\nonumber\\
		&\left.\left.+(163-101\nu)\Sp\thp+80(-1+2\nu)\Thp\Sp\right)\Sm^3\Sp\right]
		\nonumber\\
		&-\frac{\alpha_{AB}^2}{120\left(1-\zeta'\right)^2}\left\{\left[980+\nu\left(-1172+\gammaAB(-1780+1704\bar{\beta}_++320\epsilon_+)+\gammaAB^2(-421+224\bar{\beta}_+)\right.\right.\right.
		\nonumber\\
		&\left.\left.\left.+328\gammaAB^3+120\gammaAB^4+1920(\delta_-^2-\delta_+^2)+4128\bar{\beta}_++5120\bar{\beta}_+^2+1280\epsilon_+\right)\right.\right.
		\nonumber\\
		&\left.\left.+\nu^2\left(212+212\gammaAB-187\gammaAB^2-240\gammaAB^3-60\gammaAB^4+960(\delta_+^2-\delta_-^2\right)+\gammaAB\left(1364-792\bar{\beta}_+\right.\right.\right.
		\nonumber\\
		&\left.\left.\left.+320\delta_+-160\langle\epsilon\rangle\right)+\gammaAB^2\left(709+128\bar{\beta}_++80\delta_+\right)+176\gammaAB^3+20\gammaAB^4-4704\bar{\beta}_++160\bar{\beta}_+^2\right.\right.
		\nonumber\\
		&\left.\left.+320\delta_+-640\langle\epsilon\rangle\right]\Sm^2+\left[32(-163+101\nu)\thav-2560(-1+2\nu)\Thav-5216X_{AB}\thm\right.\right.
		\nonumber\\
		&\left.\left.+3232X_{AB}\nu\thm-5120X_{AB}\nu\Thm\right]\Sm+\left[2560X_{AB}\Thav-5120\nu\Thm\right]\Sp\right.
		\nonumber\\
		&\left.-\frac{32}{\left(\Sm^2-\Sp^2\right)^2}\left[\left(-80(-1+2\nu)\Thav+X_{AB}(-163+101\nu)\thm-160X_{AB}\nu\Thm\right)\Sm\Sp^4\right.\right.
		\nonumber\\
		&\left.\left.+\thav\left(40(1+\gammaAB)(1+32\nu)\Sm^2\thm+(-163+101\nu)\Sp^3\right)\Sm\Sp+80\left((-1+2\nu)\Thm\right.\right.\right.
		\nonumber\\
		&\left.\left.\left.+X_{AB}\Thp\right)\Sp^5+\left((-163+101\nu)\thm\Sp-20(1+\gammaAB)(1+32\nu)\thp^2\right)\Sm^4\right.\right.
		\nonumber\\
		&\left.\left.+\left(20(1+\gammaAB)(1+32\nu)\thm^2+(-163+101\nu)\thm\Sp+80\Sp((-1+2\nu)\Thm+X_{AB}\Thp)\right)\Sm^2\Sp^2\right.\right.
		\nonumber\\
		&\left.\left.+\left(40X_{AB}(1+\gammaAB)(1+32\nu)\thm^2+\Sp(80X_{AB}\Thm+(163-101\nu)\thp\right.\right.\right.
		\nonumber\\
		&\left.\left.\left.+80(-1+2\nu)\Thp)\right)\Sm^3\Sp\right]\right\}~,
	\nonumber\\
	\mathcal{G}^{\rm r,v^2}_{\rm 1PN} =&\left[-\frac{229}{70}+\frac{227}{30}\nu-\frac{57}{70}\nu^2+\gammaAB\left(-\frac{32}{15}+\frac{4}{15}\nu\right)+\frac{24}{5}\langle\bar{\beta}\rangle+\frac{8}{15}\nu\left(3X_{AB}\bar{\beta}_-+2\bar{\beta}_+\right)\right]\Sm^2
	\nonumber\\
	&+\frac{4}{15}\left[\left(-17-47\nu-4\gammaAB X_{AB}^2\right)\thav-2X_{AB}\left(9+31\nu\right)\thm-40\nu X_{AB}\Thm-20\left(-1+2\nu\right)\Thav\right]\Sm
	\nonumber\\
	&+\left[X_{AB}\left(\frac{286}{105}+\frac{136}{35}\nu+\frac{14}{5}\gammaAB\right)+\frac{16}{15}\left((7-14\nu)\bar{\beta}_-+X_{AB}(-7+4\nu)\bar{\beta}_+\right)\right]\Sm \Sp 
	\nonumber\\
	&+ \left[-\frac{8}{15}\left(-9+16\nu+8\nu\gammaAB\right)\thm+\frac{16}{15}\left(4+\gammaAB\right)X_{AB}\thav\right]\Sp -\left[\frac{20}{21}+\frac{76}{35}\nu\right]\Sp^2 -\frac{32}{3}X_{AB}\thm \thp
	\nonumber\\
	&+	\frac{1}{\Sm^2-\Sp^2}\left[\frac{24}{5}X_{AB}\thm \left(\Sm^3+\Sp^3\right)+\left(\frac{16}{3}(\thm^2+\thp^2)+\frac{64}{3}\nu(2\thp^2-3\thm^2\right)\Sm^2\right.
	\nonumber\\
	&\left.+\frac{24}{5}\left(-X_{AB}\thav+(-1+4\nu)\thm\right)\Sm^2\Sp-\frac{24}{5}X_{AB}\thm\Sm\Sp^2+\frac{16}{3}\left((-1+4\nu)\thm^2-\thp^2\right)\Sp^2\right.
	\nonumber\\
	&\left.+\frac{64}{3}(-1+2\nu)\thm\left(\thav+X_{AB}\thm-\thp\right)\Sm\Sp\right]~,
	\nonumber \\
\mathcal{G}^{\rm v^4}_{\rm 1PN}= &\left[\frac{69}{70}-\frac{1679}{210}\nu+\frac{131}{30}\nu^2+\gammaAB\left(\frac{2}{3}-\frac{58}{15}\nu\right)\right]\Sm^2+\left[\langle\theta\rangle\left(\frac{28}{15}+\frac{4}{5}\nu\right)-\frac{88}{15}\nu\theta_+\right]\Sm
\nonumber\\
&-X_{AB}\left[\frac{16}{15}+\frac{16}{15}\nu-\frac{14}{15}\gammaAB\right]\Sm\Sp + \left[-\frac{16}{15}X_{AB}\langle\theta\rangle+\frac{16}{5}\nu\theta_-\right]\Sp~,
\nonumber \\
\mathcal{G}^{\rm \dot{r}^4}_{\rm 1PN}= &\left[-\frac{127}{14}+\frac{106}{7}\nu+\frac{12}{7}\nu^2-\gammaAB\left(4-2\nu\right)\right]\Sm^2+\left[\langle\theta\rangle\left(12-8\nu\right)-16\nu\theta_+\right]\Sm
\nonumber\\
&-X_{AB}\left[\frac{48}{7}+\frac{86}{7}\nu+6\gammaAB\right]\Sm \Sp +\left[-8 X_{AB}\langle\theta\rangle+16\nu\theta_-\right]\Sp +\left[-\frac{11}{7}+\frac{24}{7}\nu\right]\Sp^2~,
\nonumber\\
\mathcal{G}^{\rm v^2,\dot{r}^2}_{\rm 1PN}=&\left[\frac{299}{35}-\frac{23}{5}\nu-\frac{302}{35}\nu^2+\gammaAB\left(\frac{18}{5}-\frac{16}{5}\nu\right)\right] \mathcal{S}_-^2 + \left[\frac{136}{5}\nu \theta_++ \langle\theta\rangle\left(-\frac{64}{5}+\frac{12}{5}\nu\right)\right]\Sm
\nonumber\\
&+\frac{2}{35}X_{AB}\left[128+297\nu+ 112\gammaAB\right] \Sm \Sp+ 8 \left[X_{AB}  \langle\theta\rangle- \frac{12}{5}\nu\theta_-\right]\Sp+\left[\frac{8}{7}-\frac{26}{35}\nu\right]\Sp^2~.
	\end{align}
	\end{widetext}
where the coefficient $\zeta'$ in DEF convention is 
	\begin{align}
		\zeta'=1-\alpha_{AB}\frac{(2+\gammaAB)}{2}~.
	\end{align}
	This scalar flux contribution goes to zero in GR the limit as in GR limit, we have $\alpha_{AB} \to 1$ and $\gammaAB \to 0$ implying $\zeta' \to 0$.
	
	We could not compute the explicit expression of the scalar instantaneous contribution at 1.5PN order, given that the scalar dipole moments at relative 2.5PN accuracy are not explicitly given in Ref.~\cite{Bernard:2022noq}.
	If these were to be published, it would be straightforward to substitute then into Eq.~\eqref{eq:AngMs} and compute the full 1.5PN scalar angular momentum flux.

\subsection{Tail contribution}
Following sec.~\ref{sec:tensorial-nonloc} and computation of non-local contributions to the angular momentum flux in GR~\cite{Gopakumar:1997bs, Arun:2009mc}, we compute the tail contribution to the scalar part of the angular momentum flux in ST theories up to 1.5PN order using the discrete Fourier decomposition of the scalar dipolar and quadrupolar mass moments. 
The tail contribution to the scalar part of the angular momentum flux starts at 0.5PN order, while this tail terms enter the tensorial part starting from 1.5PN order.
The tail integrals up to 1.5PN order in ST theory are given by
	\begin{align}
	\label{eq:scalar-tail}
	\mathcal{G}_i^{\rm tail, S} &= \frac{G\phi_0(1-\zeta')}{\zeta'c^3}\epsilon_{ijk} \nonumber \\
	&\times \left\{\frac{2GM}{3\phi_0c^3}{\dot{I}_j^{\rm S}}\int_{-\infty}^{T_{\rm R}}dV\overset{(4)}{I_k^{\rm S}}(V)\left[\ln\left(\frac{T_{\rm R}-V}{2\tau_0}\right)+1\right]\right.
	\nonumber \\
	&\left.-\frac{2GM}{3\phi_0c^3}{\ddot{I}_j^{\rm S}}\int_{-\infty}^{T_{\rm R}}dV\overset{(3)}{I_k^{\rm S}} (V)\left[\ln\left(\frac{T_{\rm R}-V}{2\tau_0}\right)+1\right]\right.
	\nonumber \\
	&\left. +\frac{2GM}{15\phi_0c^5}{\ddot{I}_{jl}^{\rm S}}\int_{-\infty}^{T_{\rm R}}dV\overset{(5)}{I_{kl}^{\rm S}}(V)\left[\ln\left(\frac{T_{\rm R}-V}{2\tau_0}\right)+\frac{3}{2}\right]\right.
	\nonumber\\
	&\left.-\frac{2GM}{15\phi_0c^5}\overset{(3)}{I_{jl}^{\rm S}} \int_{-\infty}^{T_{\rm R}}dV\overset{(4)}{I_{kl}}(V)\left[\ln\left(\frac{T_{\rm R}-V}{2\tau_0}\right)+\frac{3}{2}\right]
	\right\}\,.
	\end{align}
	
As shown in Ref.~\cite{Jain:2023fvt}, the scalar dipole moment at Newtonian order in polar coordinates is a periodic function of the mean anomaly $\ell$ and can be decomposed into a Fourier series, i.e. 
\begin{equation}
\label{eq:FD-Newt}
I_i^{\rm S}(\ell)= \sum_{p=-\infty}^{\infty} I_i^{\rm S} (p) e^{\mathrm{i} p \ell}
\end{equation}
where $I^{\rm S}_{i} (p)$ are discrete Fourier functions of $p$ and since the moments are real $I^{\rm S}_{i} (-p)=I^{\rm S}_{i} (p)^*$ where the superscript "$*$" denotes a complex conjugation.

We follow exactly the same method as in Ref.~\cite{Arun:2009mc} and express the tail terms in Eq.~\eqref{eq:scalar-tail} in terms of the Fourier functions of the multipole moments, averaged over the mean anomaly $\ell$. 
All of the hereditary contributions, apart from the scalar dipolar mass moment, entering Eq.~\eqref{eq:scalar-tail}  will require only relative Newtonian precision for the leading order computations, hence we can use the simple decomposition of Eq.~\eqref{eq:FD-Newt}. 
For the beyond Newtonian precision, we will follow the approach used in GR to exploit the doubly periodic nature of the dynamics and express the scalar mass dipolar moment as
\begin{equation}
\label{eqn:FD-1PN}
I^{\rm S}_{i}(\ell)= \sum_{p=-\infty}^{\infty}\sum_{m=-1}^{1} I^{\rm S}_{i} (p,m) e^{\mathrm{i} (p+km) \ell}\,,
\end{equation}
using the general formula of Ref.~\cite{Arun:2009mc}. 
Note that we here start from $m = -1$ because this is a dipolar contribution, while the GR equivalent starts from $m=-2$.

\subsubsection{Relative Newtonian Order}

In this section, we present the computation of scalar tail contribution up to 1.5PN order at the relative Newtonian precision.
We first separate Eq.~\eqref{eq:scalar-tail} into the four addends and denote the four terms with subscripts I, II, III and IV.

Let us follow the computation for $(\mathcal{G}_i^{\rm S, tail})_{\rm I}$ at Newtonian order.
We start by taking the average over the mean anomaly $\ell$, denoted as $\langle\mathcal{G}_i^{\rm S, tail} \rangle_{\rm I}$, and get
\begin{align}
\langle\mathcal{G}_i^{\rm tail, S} \rangle_{\rm I} &= \frac{G^2M(1-\zeta')}{3\zeta' \pi c^6}\epsilon_{ijk}\int_0^{2\pi} d\ell{\dot{I}_j^{\rm S}}  \nonumber \\
&\times\int_{-\infty}^{T_{\rm R}}dV{I_k^{\rm S}}^{(4)}(V)\left[\ln\left(\frac{T_{\rm R}-V}{2\tau_0}\right)+1\right]\,.
\end{align}

We can now introduce $\tau = T_R - V$.
Using the Newtonian Fourier decomposition, Eq.~\eqref{eq:FD-Newt}, and the linear dependence of $\ell$ with time, so that $\ell (t- \tau) = \ell(t)-n \tau$ (see, e.g., Ref.~\cite{Jain:2023fvt}), we obtain
\begin{align}
\langle\mathcal{G}_i^{\rm tail, S}\rangle_{\rm I} &= \mathrm{i}  \frac{2G^2M(1-\zeta')}{3\zeta' c^6}\epsilon_{ijk} \sum_{p=1}^{\infty} p^5 n^5 
	\nonumber \\
&\times \left\{I^{\rm S}_{j}(p) I^{\rm S}_{k}(p)^*\int_0^{\infty} d\tau e^{\mathrm{i} n p t}\left[\ln\left(\frac{\tau}{2\tau_0}\right)+1\right]\right.\nonumber \\
&\left.-I^{\rm S}_{k}(p) I^{\rm S}_{j}(p)^*\int_0^{\infty} d\tau e^{-\mathrm{i} n p t}\left[\ln\left(\frac{\tau}{2\tau_0}\right)+1\right]\right\}\,.
\end{align}

To solve the integral in the above equation, we use the closed-form solution of the integral
\begin{equation}
\int_0^{\infty} dt e^{\mathrm{i} \sigma t} \ln\left(\frac{t}{2t_0}\right) = -\frac{1}{\sigma}\left\{\frac{\pi}{2}\mathrm{sign}(\sigma)+\mathrm{i}\left[\ln(2|\sigma|t_0)+\gamma_{\rm E}\right]\right\}
\end{equation}
where $\mathrm{sign}$ is signum function, i.e $\mathrm{sign}(\sigma)=|\sigma|/\sigma$ and $\gamma_{\rm E}$ is Euler's constant. 
Using this integral, the equation above becomes
\begin{align}
\langle\mathcal{G}_i^{\rm S, tail} \rangle_{\rm I} =& -  
\mathrm{i} \frac{2\pi G^2M(1-\zeta')}{3\zeta'c^6}\epsilon_{ijk}  \sum_{p=1}^{\infty} p^4 n^4 I^{\rm S}_{j}(p) I^{\rm S}_{k}(p)^*\,.
\end{align}

Similarly, we can solve for the other terms in Eq.~\eqref{eq:scalar-tail},  of the tail contribution in the scalar part of the angular momentum flux, obtaining
\begin{align}
\langle\mathcal{G}_i^{\rm tail, S} \rangle_{\rm II} &=  - \mathrm{i} \frac{2\pi G^2M(1-\zeta')}{3\zeta'c^6}\epsilon_{ijk}\sum_{p=1}^{\infty} p^4 n^4 I^{\rm S}_{j}(p) I^{\rm S}_{k}(p)^*\,,\nonumber \\
\langle\mathcal{G}_i^{\rm S, tail} \rangle_{\rm III} &= -\mathrm{i}\frac{2\pi G^2M(1-\zeta')}{15\zeta'c^8}\epsilon_{ijk} \sum_{p=1}^{\infty} p^6 n^6 I^{\rm S}_{jl}(p) I^{\rm S}_{kl}(p)^*\,, \nonumber \\
\langle\mathcal{G}_i^{\rm S, tail} \rangle_{\rm IV} &=-\mathrm{i}\frac{2\pi G^2M(1-\zeta')}{15\zeta'c^8}\epsilon_{ijk} \sum_{p=1}^{\infty} p^6 n^6 I^{\rm S}_{jl}(p) I^{\rm S}_{kl}(p)^*\,.
\end{align}

We note that the dipolar contributions at Newtonian order, terms I and II above, give the same contribution. The same holds also for the Newtonian quadrupolar contributions, i.e. $\langle\mathcal{G}_i^{\rm S, tail} \rangle_{\rm III} = \langle\mathcal{G}_i^{\rm S, tail} \rangle_{\rm IV}$.

\subsubsection{First post-Newtonian order}
The leading order scalar tail contribution of Eq.~\eqref{eq:scalar-tail} requires a relative 1PN order dipolar moment as well for the scalar flux up to 1.5PN order. In this case, the Fourier decomposition of the dipolar moment is ``doubly periodic" in two variables $K \ell = (1+k)\ell$ and $\ell$, where $\ell$ is a mean anomaly and $K$ is the advance periastron per revolution and the decomposition is given in Eq.~\eqref{eqn:FD-1PN}. After inserting the Fourier decomposition, the average 1PN order scalar tail contribution reads
\begin{widetext}
\begin{align}
\label{eqn:dmoment1PN}
\langle\mathcal{G}_i^{\rm  tail, S}\rangle =&- \mathrm{i}\frac{2G^2M(1-\zeta')}{3\zeta'c^6}\epsilon_{ijk}\sum_{p,p';m,m'}(p+mk)(p'+m'k)^3[p'-p+(m'-m)k]I_j^{\rm S}(p,m)I_k^{\rm S}(p',m')\nonumber\\
&\times \langle e^{\mathrm{i}(p+p'+mk+m'k)l}\rangle\int^{\infty}_0d\tau e^{-\mathrm{i}(p'+m'k) n\tau} \left[\ln\left(\frac{\tau}{2\tau_0}\right)+1\right]~,
\end{align}
\end{widetext}
where the summation over $p$ and $p'$ ranges from $-\infty$ to $\infty$, the summation over $m$ and $m'$ ranges from -1 to 1 and the angular brackets denote an average over $\ell$. 
The explicit Fourier decomposition up to 1PN order can be obtained following Ref.~\cite{Arun:2009mc}, which we leave for future work.

	\section{Discussion}

In this work, we present the angular momentum flux emitted by non-circular nonspinning binaries in massless ST theories up to 1.5PN order.
We build upon the results of Refs.~\cite{Lange:2018pyp, Lange:2017wki, Bernard:2022noq} and compute, for the first time, tensorial contributions to the angular momentum flux at 1.5PN order and scalar contributions up to 1PN order.
We only write implicit expressions for the 1.5PN instantaneous scalar angular-momentum flux, because the scalar dipole moments at relative 2.5PN accuracy are not explicitly written in Ref.~\cite{Bernard:2022noq}.

For the tensorial contributions, we first find ST deformations to the instantaneous GR angular momentum flux, Eqs.~\eqref{eq:GT}--\eqref{eq:tens-inst15pn}.
Due to the scalar dipolar radiation present in ST theories, we also find that the memory effects in the tensorial flux start 1PN order before their GR counterparts, while tail contributions start at the same (1.5PN) order and coincide with GR computations. 
To compute the leading order contribution to the non-zero integral of memory effects, we use the Keplerian motion in ST theories and obtain a pure zero-frequency component of the memory (see Sec.~\ref{sec:tensorial-nonloc}).

Next, to derive the scalar contribution to the angular momentum flux, we first derive the relation of the scalar angular momentum flux to the radiative scalar multipole moments. 
Then, using this relation, we first compute the instantaneous contribution up to 1PN order [Eqs.~\eqref{eq:G-1PN}--\eqref{eq:GS}].
We then compute the hereditary effects to the scalar part in the angular momentum flux. 
As there is a dipolar radiation due to the scalar field, we find that both the instantaneous and the hereditary effects in the scalar part start at 1PN order less relative to the leading order quadrupolar term in GR, i.e. the hereditary contribution starts at 0.5 PN order and instantaneous contribution starts at -1PN order relative to GR. 
To find the closed form expression for the hereditary contribution, we use the discrete Fourier decomposition of the scalar radiative multipole moments (the Fourier decomposition of dipolar multipole moment is given in App.~A of Ref.~\cite{Jain:2023fvt} and the Fourier decomposition of quadrupolar multipole moment follows the Fourier decomposition of GR multipole moment). 

As a first consistency test of our results, we check that the angular momentum flux coincides with the GR angular momentum flux in the GR limit, when all ST parameters reduce to zero.
Second, we study our results under the assumption that the sensitivity ($s_I$) for the isolated black holes holds for the binary black hole systems as well. 
For isolated black holes $s_I = 1/2$ (see, Ref.~\cite{Hawking:1972hy}), which relate to our conventions as
\begin{align*}
s_I = \frac{1}{2}-\frac{\alpha^0_I}{2\alpha_0}~.
\end{align*}
The isolated black hole scenario implies that $\alpha_I^0 = 0$, and hence also $\bar{\beta}_I = \epsilon_I = \delta_I = 0$. 
This would entail that our results are indistinguishable from the GR ones for binary black hole systems, i.e. the angular momentum flux matches with the GR angular momentum flux of Refs.~\cite{Peters:1964zz, Junker:1992kle,Gopakumar:1997bs, Arun:2009mc}.
We also add that, while this paper was in preparation, we became aware that a similar computation was being performed~\cite{Trestini:inprep}. 
We compared our results for the general expression of the instantaneous part of the angular momentum flux at 1PN, Eq.~\eqref{eq:GS}, and found perfect agreement.

These results should be considered as a step forward to construct waveform templates for generic orbits in ST theories. 
In future work, we will use the results obtained here for the angular momentum flux and expressions of energy flux from Ref.~\cite{Bernard:2022noq} to obtain the radiation-reaction force in effective-one-body formalism for \textit{generic} orbits.

\section*{Acknowledgments}

The authors are grateful to David Trestini for insightful discussions and cross-checking of some computations. T.~J.~also thanks Ulrich Sperhake for useful discussions. 
T.~J. and P.~R. thank the hospitality and the stimulating environment of the Institut des Hautes Etudes Scientifiques.  
T.~J. is jointly funded by Cambridge Trust, the Department of Applied Mathematics and Theoretical Physics (DAMTP), and the Centre for Doctoral Training, University of Cambridge.
P.~R. is supported by the Italian Minister of University and Research (MUR) via the 
PRIN 2020KB33TP, {\it Multimessenger astronomy in the Einstein Telescope Era (METE)}.
The present research was also partly supported by the ``\textit{2021 Balzan Prize for 
Gravitation: Physical and Astrophysical Aspects}'', awarded to Thibault Damour. 
	
\appendix	
\section{Einstein and Jordan-frame parameters}
\label{app:st_pars}
Here we collect the conversion of the Jordan-Frame parameters of Refs.~\cite{Bernard:2022noq} to DEF conventions \cite{Damour:1992we},\ i.e. the Einstein-Frame parameters. These parameters are translated using \eqref{eq:alpha}, \eqref{eq:Jordan_g} and \eqref{eq:jordan_m}, and are gathered in Table \ref{LauravsDEF}.

\begin{table}[h]
\begin{ruledtabular}
\caption{\label{LauravsDEF} Conversion of two-body parameters}
   \centering
   \renewcommand{\arraystretch}{1.7}
\begin{tabular}{c l c}
   LB\cite{Bernard:2018hta} & DEF~\cite{Damour:1992we,Damour:1995kt} & This paper \\
  \hline
  $m_{1}$ & $m^0_{A}/\mathcal A_0$ & $ m^0_{A}/\mathcal A_0 \equiv \tilde{m}^0_{A}$ \\
  $m_{2}$ & $m^0_{B}/\mathcal A_0$ & $ m^0_{B}/\mathcal A_0 \equiv \tilde{m}^0_{B}$ \\
  $\alpha$ & $\frac{1+\alpha^0_A\alpha^0_B}{1+\alpha_0^2}$ & $\alpha_{AB}$\\
  $\tilde{G}\alpha$ & $(1+\alpha^0_A\alpha^0_B)\mathcal A^2_0\equiv G_{AB}\mathcal A_0^2$ & $G_{AB}\mathcal A_0^2\equiv\tilde{G}_{AB}$ \\
  $\bar{\gamma}$ & $-2\frac{\alpha^0_A\alpha^0_B}{1+\alpha^0_A\alpha^0_B}\equiv\bar{\gamma}_{AB}$ & $\bar{\gamma}_{AB}$ \\
  $\bar{\beta}_1$ & $\frac{1}{2}\frac{(\beta_A\alpha_B^2)_0}{(1+\alpha_A^0\alpha_B^0)^2}\equiv\bar{\beta}^A_{BB}$ & $\bar{\beta}_A$\\
  $\bar{\beta}_2$ & $\frac{1}{2}\frac{(\beta_B\alpha_A^2)_0}{(1+\alpha_A^0\alpha_B^0)^2}\equiv\bar{\beta}^B_{AA}$ & $\bar{\beta}_B$\\
  $\bar{\delta}_1$ & $\frac{(\alpha_A^0)^2}{(1+\alpha_A^0\alpha_B^0)^2}$  & $\delta_A$ \\
  $\bar{\delta}_2$ & $\frac{(\alpha_B^0)^2}{(1+\alpha_A^0\alpha_B^0)^2}$  & $\delta_B$\\
  $\bar{\chi}_1$ & $-\frac{1}{4}\frac{(\beta'_A\alpha_B^3)_0}{(1+\alpha_A^0\alpha_B^0)^3}\equiv-\frac{1}{4}\epsilon^A_{BBB}$ & $-\frac{1}{4}\epsilon_A$\\
  $\bar{\chi}_2$ & $-\frac{1}{4}\frac{(\beta'_B\alpha_A^3)_0}{(1+\alpha_A^0\alpha_B^0)^3}\equiv-\frac{1}{4}\epsilon^B_{AAA}$ & $-\frac{1}{4}\epsilon_B$
\end{tabular}
\end{ruledtabular}
\end{table}

	\bibliographystyle{apsrev4-2}
	\bibliography{refs.bib, local.bib}
	
\end{document}